\documentclass[twocolumn,nofootinbib,floatfix,amsmath,amssymb,showpacs,superscriptaddress]{revtex4}

\usepackage{graphicx}
\usepackage{dcolumn}
\usepackage{bm}
\usepackage{color}

\begin{document}

\title{Pseudoscalar-meson--Octet-baryon Coupling Constants in Two-flavor Lattice QCD}

\author{G\"{u}ray Erkol}
\affiliation{Ozyegin University, Kusbakisi Caddesi No:2 Altunizade, Uskudar Istanbul 34662 Turkey}
\affiliation{Department of Physics, H-27, Tokyo Institute of Technology, Meguro, Tokyo 152-8551 Japan}
\author{Makoto Oka}%
\affiliation{Department of Physics, H-27, Tokyo Institute of Technology, Meguro, Tokyo 152-8551 Japan}
\author{Toru T. Takahashi}
\affiliation{Yukawa Institute for Theoretical Physics, Kyoto University,
Sakyo, Kyoto 606-8502, Japan}

\date{\today}

\begin{abstract}
We evaluate the $\pi N\!N$, $\pi\Sigma\Sigma$, $\pi\Lambda\Sigma$, $K\Lambda N$ and $K \Sigma N $ coupling constants and the corresponding monopole masses in lattice QCD with two flavors of dynamical quarks. The parameters representing the SU(3)-flavor symmetry are computed at the point where the three quark flavors are degenerate at the physical $s$-quark mass. In particular, we obtain $\alpha\equiv F/(F+D)=0.395(6)$. The quark-mass dependences of the coupling constants are obtained by changing the $u$- and the $d$-quark masses. We find that the SU(3)-flavor parameters have weak quark-mass dependence and thus the SU(3)-flavor symmetry is broken by only a few percent at each quark-mass point we consider.

\end{abstract}
\maketitle
Meson-baryon coupling constants are important ingredients for hadron physics as they provide a measure of baryon-baryon interactions in terms of One Boson Exchange (OBE) models, and production of mesons off the baryons. In phenomenological potential models, the meson-baryon coupling constants are determined so as to reproduce the nucleon-nucleon, hyperon-nucleon and the hyperon-hyperon interactions in terms of, {\it e.g.}, OBE models. On the other hand, it is an important issue to determine the coupling constants at the hadronic vertices directly from QCD, the underlying theory of the strong interactions. The only method we know that provides a first-principles calculation of hadronic phenomena is lattice QCD, which serves as a valuable tool to determine the hadron couplings in a model-independent way.

Among other meson-baryon coupling constants, the $\pi N\!N$ coupling constant, $g_{\pi N\!N}$, which enters as a fundamental quantity in low-energy dynamics of nucleon-nucleon and pion-nucleon, has been a subject of intense investigation. It is defined as the $\pi N\!N$ form factor, $g_{\pi N\!N}(q^2)$, at zero momentum transfer, $q^2=0$. The value of the coupling constant at the pion pole is relatively well-known from experiment: $g_{\pi N\!N}^2(m_\pi^2)/4\pi\simeq 13.6$ (see, {\it e.g.}, Ref.~\cite{Ericson:2000md, Bugg:2004cm} for a review). The value at zero momentum transfer can be extracted from the Goldberger-Treimann relation (GTR), $g_{\pi N\!N}\equiv g_A m_N/f_\pi\sim 12.8$, where $f_\pi$ is the pion decay constant and $m_N$ and $g_A$ are the mass and the axial-vector coupling constant of the nucleon, respectively. An earlier determination of the $\pi N\!N$ coupling constant from a quenched-lattice QCD calculation, which reports $g_{\pi N\!N}=12.7 \pm 2.4$~\cite{Liu:1994dr}, is in agreement with the phenomenological value. In the SU(3)-flavor~[SU(3)$_F$] symmetric limit, one can classify the pseudoscalar-meson--octet-baryon coupling constants in terms of two parameters: the $\pi N\!N$ coupling constant and the $\alpha=F/(F+D)$ ratio of the pseudoscalar octet~\cite{deSwart:1963gc}. This systematic classification is expected to govern all the meson-baryon couplings, however as we move from the symmetric case to the realistic one, the SU(3)$_F$ breaking occurs as a result of the $s$-quark mass and the physical masses of the baryons and mesons. The broken symmetry no longer provides a pattern for the meson-baryon coupling constants, and therefore they should be individually calculated based on the underlying theory, QCD.

In this work we extract the coupling constants $g_{\pi N\!N}$, $g_{\pi\Sigma\Sigma}$, $g_{\pi\Lambda\Sigma}$, $g_{K\Lambda N }$ and $g_{K\Sigma N}$ (denoted by $g_{MBB^\prime}$ hereafter) by employing lattice QCD with two flavors of dynamical quarks. The evaluation of the coupling constants allows us to check the following SU(3)$_F$ relations:
\begin{align}
\begin{split}
		g_{\pi N\!N}=g,\quad g_{\pi\Sigma\Sigma}=2 g \alpha, \quad g_{\pi\Lambda\Sigma}= \frac{2}{\sqrt{3}}g(1-\alpha),\\
		g_{K\Lambda N}=-\frac{1}{\sqrt{3}}g(1+2\alpha),\quad g_{K\Sigma N}=g(1-2\alpha),
\label{su3rel}
\end{split}
\end{align}%
which phenomenologically work rather well but are not known {\it a priori} to hold. 

\begin{figure}[th]
	\includegraphics[scale=0.60]{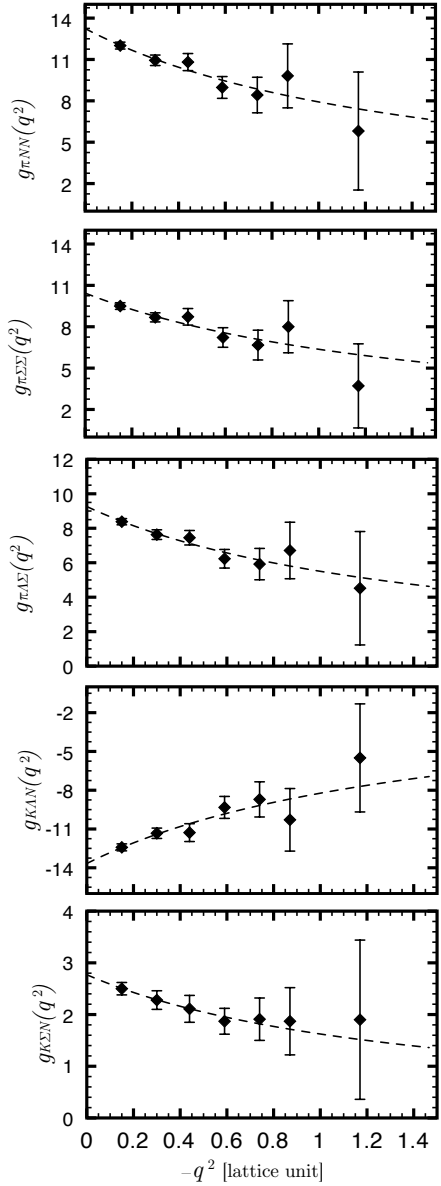}
	\caption{\label{q2dep} The $q^2$ dependence of the form factors, $g_{M B B^\prime}$ for $\kappa_{val}^{u,d}=0.1393$. The diamonds show the lattice data, and the solid curves denote the fitted form factors.}
\end{figure}	

\begin{figure}[th]
	\includegraphics[scale=0.40]{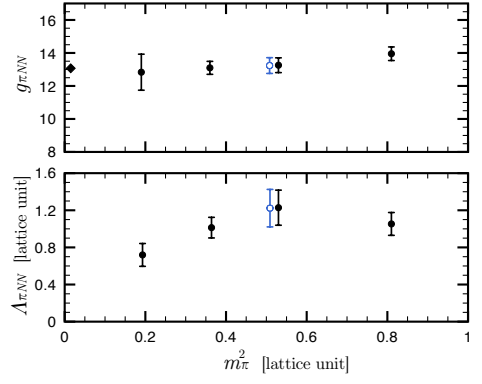}
	\caption{\label{piNN_sm}$g_{\pi N\!N}$ and $\Lambda_{\pi N\!N}$ as a function of $m_\pi^2$. The empty circle denotes the SU(3)$_F$ limit and the diamond marks the experimental result. 
}
\end{figure}	

\begin{figure}[th]
	\includegraphics[scale=0.40]{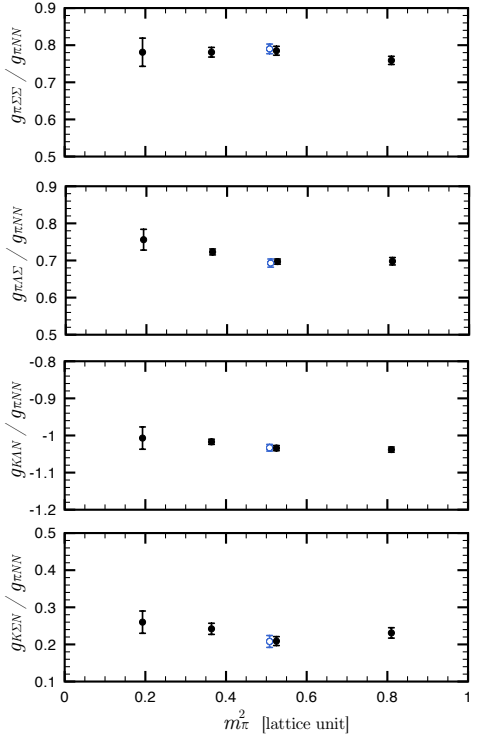}
	\caption{\label{rat_coup_sm} The $\pi\Sigma\Sigma$, $\pi\Lambda\Sigma$, $K\Lambda N$ and $K\Sigma N$ coupling constants normalized with $g_{\pi N\!N}$ as a function of $m_\pi^2$. The empty circle denotes the SU(3)$_F$ limit. 
}
\end{figure}

The pseudoscalar current matrix element is written as
\begin{equation}
	\langle {\cal B}({\bf p}) |P(0)|{\cal B}^\prime({\bf p}^\prime) \rangle = g_P(q^2)\bar{u}({\bf p}) i \gamma_5 u({\bf p}^\prime),
\end{equation}
where $g_P(q^2)$ is the pseudoscalar form factor, $q_\mu=p_\mu^\prime-p_\mu$ is the transferred four-momentum and $P(x)=\bar{\psi}(x)i\gamma_5 \frac{\tau_3}{2}\psi(x)$ is the pseudoscalar current. We compute this matrix element using the ratio~\cite{Alexandrou:2006mc,Alexandrou:2007zz}
\begin{align}
\begin{split}\label{ratio}
	&R(t_2,t_1;{\bf p}^\prime,{\bf p};\Gamma;\mu)=\\
	&\quad\frac{\langle G^{{\cal B} {\cal P} {\cal B}^\prime}(t_2,t_1; {\bf p}^\prime, {\bf p};\Gamma)\rangle}{\langle G^{{\cal B}^\prime}(t_2; {\bf p}^\prime;\Gamma_4)\rangle} \left[\frac{\langle G^{{\cal B}}(t_2-t_1; {\bf p};\Gamma_4)\rangle}{\langle G^{{\cal B}^\prime}(t_2-t_1; {\bf p}^\prime;\Gamma_4)\rangle}\right.\\ 
	&\quad\left.\times\frac{\langle G^{{\cal B}^\prime}(t_1; {\bf p}^\prime;\Gamma_4)\rangle \langle G^{{\cal B}^\prime}(t_2; {\bf p}^\prime;\Gamma_4)\rangle}{\langle G^{{\cal B}}(t_1; {\bf p};\Gamma_4)\rangle \langle G^{{\cal B}}(t_2; {\bf p};\Gamma_4)\rangle} \right]^{1/2},
\end{split}
\end{align}
where the baryonic two- and three-point correlation functions are respectively defined as
\allowdisplaybreaks{
\begin{align}
	\begin{split}\label{twopcf}
	&\langle G^{{\cal B}}(t; {\bf p};\Gamma_4)\rangle=\sum_{\bf x}e^{-i{\bf p}\cdot {\bf x}}\Gamma_4^{\alpha\alpha^\prime} \\
	&\qquad\times \langle \text{vac} | T [\eta_{\cal B}^\alpha(x) \bar{\eta}_{{\cal B}^\prime}^{\alpha^\prime}(0)] | \text{vac}\rangle,
	\end{split}\\
	\begin{split}
	&\langle G^{{\cal B P B^\prime}}(t_2,t_1; {\bf p}^\prime, {\bf p};\Gamma)\rangle=-i\sum_{{\bf x_2},{\bf x_1}} e^{-i{\bf p}\cdot {\bf x_2}} e^{i{\bf q}\cdot {\bf x_1}} \\
	&\qquad\times \Gamma^{\alpha\alpha^\prime} \langle \text{vac} | T [\eta_{\cal B}^\alpha(x_2) P(x_1) \bar{\eta}_{{\cal B}^\prime}^{\alpha^\prime}(0)] | \text{vac}\rangle,
	\end{split}
\end{align}
}%
with $\Gamma\equiv\gamma_3 \gamma_5 \Gamma_4$ and $\Gamma_4\equiv (1+\gamma_4)/2$. The baryon interpolating fields are given as
\begin{align}
	\begin{split}\raisetag{60pt}
		\eta_N(x)&=\epsilon^{abc}[u^{T a}(x) C \gamma_5 d^b(x)]u^c(x),\\
		\eta_\Sigma(x)&=\epsilon^{abc}[s^{T a}(x) C \gamma_5 u^b(x)]u^c(x),\\
		\eta_\Lambda(x)&=\frac{1}{\sqrt{6}}\epsilon^{abc}\{[u^{T a}(x) C \gamma_5 s^b(x)]d^c(x)-[d^{T a}(x)C\\
		&\quad \times \gamma_5 s^b(x)]u^c(x)+2[u^{T a}(x) C \gamma_5 d^b(x)]s^c(x)\},
	\end{split}
\end{align}
where $C=\gamma_4\gamma_2$ and $a$, $b$, $c$ are the color indices. $t_1$ is the time when the meson interacts with a quark and $t_2$ is the time when the final baryon state is annihilated. The ratio in Eq.~(\ref{ratio}) reduces to the desired pseudoscalar form factor when $t_2-t_1$ and $t_1\gg a$, {\it viz.}
\begin{equation}\label{desratio}
	R(t_2,t_1;{\bf 0},{\bf p};\Gamma;\mu)\xrightarrow[t_2-t_1\gg a]{t_1\gg a} \frac{g_P^L(q^2)}{[2 E (E+m)]^{1/2}} \, q_3,
\end{equation}
where $m$ and $E$ are the mass and the energy of the initial baryon, respectively, and $g_P^L(q^2)$ is the lattice pseudoscalar form factor. Since the ratio in \eqref{desratio} is proportional to the transfered momentum $q_3$, it cannot be used directly to obtain $g_P^L(q^2)$ at $q^2=0$. We apply a procedure (similarly to the one in Ref.~\cite{Liu:1994dr}) of seeking plateau regions as a function of $t_1$ in the ratio \eqref{desratio} and calculating $g_P^L(q^2)$ at the momentum transfers ${\bf q}^2 a^2=n (2\pi/L)^2$ (for the lowest nine $n$ points), where $L$ is the spatial extent of the lattice. We then obtain the meson-baryon form factor via the relation
\begin{equation}
	g_P^L(q^2)=\frac{G_M\,g_{M B B^\prime}(q^2)}{m_M^2-q^2},
\label{MBff}
\end{equation}
assuming that the pseudoscalar form factors are dominated by the pseudoscalar-meson poles. Here $G_M\equiv\langle \text{vac} \lvert P(0) \rvert M \rangle$ is extracted from the two-point mesonic correlator $\langle P(x)P(0)\rangle$. Finally we extract the meson-baryon coupling constants $g_{M B B^\prime}=g_{M B B^\prime}(0)$ by means of a monopole form factor:
\begin{equation}
	g_{M B B^\prime}(q^2)=g_{M B B^\prime} \frac{\Lambda^2_{M B B^\prime}}{\Lambda^2_{M B B^\prime}-q^2}.	
\end{equation}

We employ a $16^3\times 32$ lattice with two flavors of dynamical quarks and use the gauge configurations generated by the CP-PACS collaboration~\cite{AliKhan:2001tx} with the renormalization group improved gauge action and the mean-field improved clover quark action. We use the gauge configurations at $\beta=1.95$ with the clover coefficient $c_{SW}=1.530$, which give a lattice spacing of $a=0.1555(17)$ fm ($a^{-1}=1.267$ GeV), which is determined from the $\rho$-meson mass. The simulations are carried out with four different hopping parameters for the sea and the $u$,$d$ valence quarks, $\kappa_{sea},\kappa_{val}^{u,d}=$ 0.1375, 0.1390, 0.1400 and 0.1410, which correspond to quark masses of $\sim$ 150, 100, 65, and 35~MeV, and we use 490, 680, 680 and 490 such gauge configurations, respectively. The hopping parameter for the $s$ valence quark is fixed to $\kappa_{val}^{s}=0.1393$ so that the Kaon mass is reproduced~\cite{AliKhan:2001tx}, which corresponds to a quark mass of $\sim 90$~MeV. We employ smeared source and smeared sink, which are separated by 8 lattice units in the temporal direction. Source and sink operators are smeared in a gauge-invariant manner with the root mean square radius of 0.6 fm. All the statistical errors are estimated via the jackknife analysis.

\begin{figure}[th]
	\includegraphics[scale=0.40]{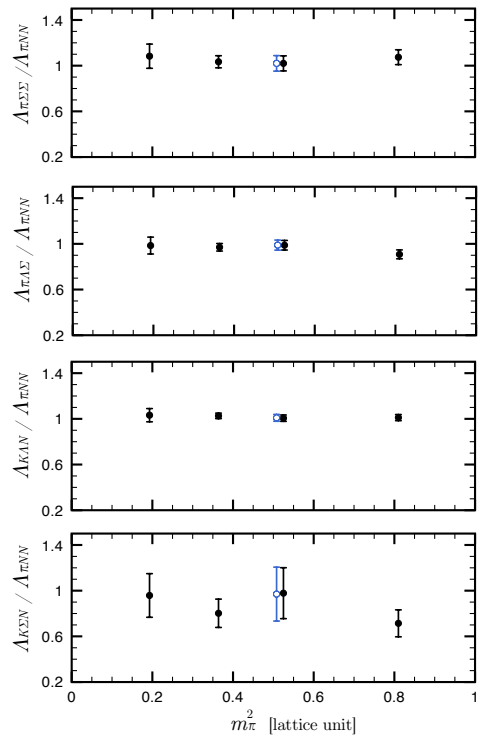}
	\caption{\label{rat_lambda} Same as Fig.~\ref{rat_coup_sm} but for monopole masses $\Lambda_{\pi\Sigma\Sigma}$, $\Lambda_{\pi\Lambda\Sigma}$, $\Lambda_{K\Lambda N}$ and $\Lambda_{K\Sigma N}$ normalized with $\Lambda_{\pi N\!N}$.}
\end{figure}

\begin{table}[th]
	\caption{The fitted values of $m_\pi$, $m_K$, $m_N$, $m_\Lambda$ and $m_\Sigma$ in lattice units. 
}
\begin{center}
\begin{tabular}{cccccc}
		\hline\hline 
		$\kappa^{u,d}_{val}$ & $m_\pi$ & $m_K$ & $m_N$ & $m_\Lambda$ & $m_\Sigma$  \\[0.5ex]
		\hline 
		0.1375 & 0.899(1)& 0.834(1)& 1.707(06)& 1.658(06)& 1.648(06)\\
		0.1390 & 0.737(1)& 0.725(1)& 1.475(05)& 1.466(06)& 1.464(06)\\
		0.1393 & 0.713(1)& 0.713(1)& 1.455(06)& 1.455(06)& 1.455(06)\\
		0.1400 & 0.603(1)& 0.635(1)& 1.289(05)& 1.312(04)& 1.318(05)\\
		0.1410 & 0.440(1)& 0.533(1)& 1.051(08)& 1.114(06)& 1.134(07)\\
		\hline\hline
\end{tabular}
	\label{mass_table}
\end{center}
\end{table}
\begin{table*}[ht]
	\caption{The fitted value of the $\pi N\!N$ coupling constant and the corresponding monopole mass (given in lattice units), together with the fitted values of the $\pi\Sigma\Sigma$, $\pi\Lambda\Sigma$, $K\Lambda N$ and $K\Sigma N$ coupling constants and the corresponding monopole masses normalized with $g_{\pi N\!N}$ and $\Lambda_{\pi N\!N}$, respectively. Here, we define $g^R_{M B\!B^\prime}=g_{M B\!B^\prime}/g_{\pi N\!N}$ and $\Lambda^R_{M B\!B^\prime}=\Lambda_{M B\!B^\prime}/\Lambda_{\pi N\!N}$. 
}
\begin{center}
\begin{tabular*}{1.0\textwidth}{@{\extracolsep{\fill}}ccccccccccc}
		\hline\hline 
		$\kappa^{u,d}_{val}$ & $g_{\pi N\!N}$ & $\Lambda_{\pi N\!N}$ & $g^R_{\pi \Sigma\Sigma}$ & $g^R_{\pi\Lambda\Sigma}$ & $g^R_{K\Lambda N}$ & $g^R_{K\Sigma N}$ & $\Lambda^R_{\pi\Sigma\Sigma}$ & $\Lambda^R_{\pi\Lambda\Sigma}$ & $\Lambda^R_{K\Lambda N}$ & $\Lambda^R_{K\Sigma N}$ \\[0.5ex]
		\hline 
		0.1375 & 13.953(412)& 1.053(123)& 0.759(11)& 0.698(11)& -1.038(07)& 0.231(14)& 1.074(065)& 0.908(039)& 1.011(27)& 0.714(118)\\
		0.1390 & 13.257(448)& 1.228(189)& 0.785(12)& 0.697(07)& -1.034(07)& 0.209(12)& 1.020(066)& 0.988(042)& 1.006(28)& 0.978(223)\\
		0.1393 & 13.236(478)& 1.223(202)& 0.789(13)& 0.699(08)& -1.033(08)& 0.209(13)& 1.020(068)& 0.989(044)& 1.009(30)& 0.970(236)\\
		0.1400 & 13.098(393)& 1.013(111)& 0.781(13)& 0.723(08)& -1.017(07)& 0.242(15)& 1.034(053)& 0.970(033)& 1.026(24)& 0.802(124)\\
		0.1410 & 12.834(1.092)& 0.719(123)& 0.781(38)& 0.756(28)& -1.007(30)& 0.260(30)& 1.083(106)& 0.985(074)& 1.032(58)& 0.958(191)\\
		\hline\hline
\end{tabular*}
	\label{res_table}
\end{center}
\end{table*}
In Table~\ref{mass_table}, we give the fitted values of the meson and baryon masses as obtained from the two-point correlation function in Eq.~\eqref{twopcf}. We extract the meson-baryon coupling constants, $g_{M B B^\prime}$, and the corresponding monopole masses, $\Lambda_{M B B^\prime}$, for each $\kappa_{val}^{u,d}$. In Fig.~\ref{q2dep}, the $q^2$ dependence of the form factors, $g_{M B B^\prime}$, for $\kappa_{val}^{u,d}=0.1393$ is given. Our complete results are presented in Table~\ref{res_table}: We give the fitted value of the $\pi N\!N$ coupling constant and the corresponding monopole mass in lattice unit, as well as the fitted values of the $\pi\Sigma\Sigma$, $\pi\Lambda\Sigma$, $K\Lambda N$ and $K\Sigma N$ coupling constants and the corresponding monopole masses normalized with $g_{\pi N\!N}$ and $\Lambda_{\pi N\!N}$, respectively. In Table~\ref{res_table}, $g^R_{M B\!B^\prime}$ and $\Lambda^R_{M B\!B^\prime}$ denote $g_{M B\!B^\prime}/g_{\pi N\!N}$ and $\Lambda_{M B\!B^\prime}/\Lambda_{\pi N\!N}$, respectively. We expect that the systematic errors cancel out to some degree in the ratios of the coupling constants and those of the monopole masses. We give a graphical representation of our results in Figs.~\ref{piNN_sm},~\ref{rat_coup_sm} and \ref{rat_lambda}. In Fig.~\ref{piNN_sm} we plot $g_{\pi N\!N}$ and $\Lambda_{\pi N\!N}$ as a function of the pion-mass squared. The ratios of the $\pi\Sigma\Sigma$, $\pi\Lambda\Sigma$, $K\Lambda N$ and $K\Sigma N$ coupling constants to the $\pi N\!N$ coupling constant, and the corresponding monopole masses normalized with $\Lambda_{\pi N\!N}$ are shown in Fig.~\ref{rat_coup_sm}. $g_{\pi N\!N}$ is consistent with the experimental value at $\kappa \le 0.1393$ and $\Lambda_{\pi N\!N}$ decreases towards the chiral limit. Note that, in addition to the monopole form, we have tried fitting the form factors to dipole and exponential forms, which have produced coupling-constant ratios consistent with those given in Table~\ref{res_table}. 

\begin{figure}[bth]
	\includegraphics[scale=0.55]{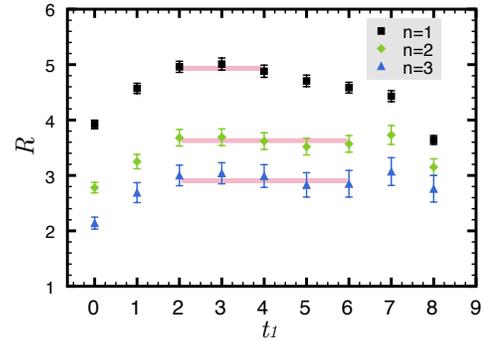}
	\caption{\label{plateau} The ratio in Eq.~\eqref{ratio} for $g_{\pi N\!N}$ as a function of current-insertion point, $t_1$, at $\kappa_{val}^{u,d,s}=0.1393$ for the first three momentum-transfer values. The bands represent the adopted plateau regions.}
\end{figure}

Having discussed the results for $g_{\pi N\!N}$, we proceed with the octet-meson--baryon coupling constants. We first concentrate on the SU(3)-flavor symmetric case, where $\kappa_{val}^{u,d}\equiv\kappa_{val}^s = 0.1393$ and the SU(3)$_F$ relations in Eq.(\ref{su3rel}) are exact. (Here we take $\kappa_{sea}^{u,d}=$0.1390 and neglect the difference in the sea-quark effects.) As expected, all the coupling ratios, $g^R_{\pi\Sigma\Sigma}$, $g^R_{\pi\Lambda\Sigma}$, $g^R_{K\Lambda N}$, and $g^R_{K\Sigma N}$ are well reproduced with $\alpha=0.395(6)$, which is obtained by a global fit. The ratios of the monopole masses, $\Lambda^R_{\pi\Sigma\Sigma}$, $\Lambda^R_{\pi\Lambda\Sigma}$, $\Lambda^R_{K\Lambda N}$, and $\Lambda^R_{K\Sigma N}$, are consistent with unity. The obtained value of $\alpha$ is consistent with that in the SU(6) spin-flavor symmetry ($\alpha=2/5$)~\cite{Pais:1966}, which is the symmetry based on the nonrelativistic quark model. We have also tried fixing all the quark masses at $\kappa_{val}^{u,d,s}=0.1390$. $g_{\pi N\!N}$ and the ratios of the coupling constants obtained in this case are as follows: $g_{\pi N\!N}=12.769(495)$, $g^R_{\pi\Sigma\Sigma}=0.785(10)$, $g^R_{\pi\Lambda\Sigma}=0.704(6)$, $g^R_{K\Lambda N}=-1.003(6)$, and $g^R_{K\Sigma N}=0.211(10)$. We have found that the coupling constants again satisfy SU(3)$_F$ and the resulting $\alpha=0.387(5)$ is consistent with that obtained at $\kappa_{val}^{u,d,s}=0.1393$. In Fig.~\ref{plateau} we plot the ratio in Eq.~\eqref{ratio} for $g_{\pi N\!N}$ as a function of current-insertion point, in order to show the plateau regions. We present the data at $\kappa_{val}^{u,d,s}=0.1393$ for the first three momentum transfer values. 

\begin{figure}[th]
	\includegraphics[scale=0.40]{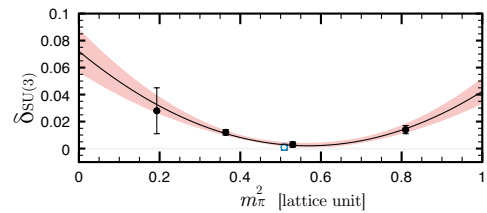}
	\caption{\label{su3brfig} The value of $\delta_{SU(3)}$ as a function of $m_\pi^2$. The curve and the shaded region denote linear chiral extrapolations with errors.}
\end{figure}
\begin{figure}[th]
	\includegraphics[scale=0.40]{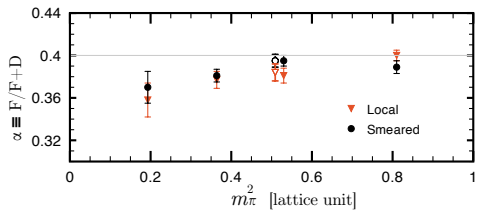}
	\caption{\label{alpha_fig} The value of $\alpha$ as obtained from a global fit of the SU(3)$_F$ relations at each quark mass we consider (in black filled circles). We also show our results as obtained with local source and local sink (in red triangles). The empty circle and the triangle denote the SU(3)$_F$ limit. The line at $\alpha=0.4$ is shown for reference only.}
\end{figure}

We next discuss the SU(3)$_F$ broken case. The quark-mass dependences we find for $g^R_{MBB'}$ and $\Lambda^R_{MBB'}$ are not large. The ratios of the coupling constants, $g^R_{MBB'}$, are similar in value to those in the SU(3)$_F$ symmetric limit, and the monopole-mass ratios, $\Lambda^R_{MBB'}$, are almost unity independently of the quark masses. This suggests that the SU(3)$_F$ breaking is small at the quark masses we consider. Our data do not allow a direct determination of SU(3)$_F$ breaking in the chiral limit, as we have flavor-symmetric data only at $\kappa_{val}^{u,d}\equiv\kappa_{val}^s = 0.1393$. On the other hand, the value of $g_{\pi N\!N}$ at the chiral point is well-known, which may serve as a reference point for us to obtain a measure of SU(3)$_F$ breaking. For this purpose, we construct the following three sets of relations:
\begin{align}\label{su3br}
	\begin{split}\raisetag{30pt}
	A_1\equiv &\frac{1}{2}\left(\sqrt{3} g^R_{\pi\Lambda\Sigma} + g^R_{\pi\Sigma\Sigma}\right),\quad A_2\equiv g^R_{K\Sigma N} + g^R_{\pi\Sigma\Sigma},\\
	A_3\equiv &\frac{1}{2} \left(g^R_{K\Sigma N}-\sqrt{3} g^R_{K\Lambda N}\right),\quad A_4\equiv -g^R_{\pi\Sigma \Sigma} - \sqrt{3} g^R_{K\Lambda N},	\\
	A_5\equiv &\frac{1}{\sqrt{3}} \left(g^R_{\pi\Lambda \Sigma} - g^R_{K\Lambda N}\right),\quad A_6\equiv \sqrt{3} g^R_{\pi\Lambda \Sigma} - g^R_{K\Sigma N},
	\end{split}\\
	\begin{split}
	&B_1\equiv \frac{1}{4}\left (\sqrt{3} g^R_{\pi\Lambda\Sigma}+3 g^R_{\pi\Sigma\Sigma}+2 g^R_{K\Sigma N}\right),\\
	&B_2\equiv \frac{1}{4}\left (2 g^R_{\pi\Sigma\Sigma}+3 g^R_{K\Sigma N}-\sqrt{3} g^R_{K\Lambda N}\right),\\
	&B_3\equiv \frac{1}{\sqrt{12}}\left (g^R_{\pi\Lambda\Sigma}-4 g^R_{K\Lambda N}-\sqrt{3} g^R_{\pi\Sigma \Sigma}\right),\\
	&B_4\equiv \frac{1}{\sqrt{12}}\left (4 g^R_{\pi\Lambda\Sigma}-\sqrt{3} g^R_{K\Sigma N}-g^R_{K\Lambda N}\right),
	\end{split}\\
	\intertext{and}
	\begin{split}\label{su3br3}
	C_1\equiv &\frac{1}{2}\left (\sqrt{3} g^R_{\pi\Lambda\Sigma}-\sqrt{3} g^R_{K\Lambda N}-g^R_{\pi\Sigma \Sigma}-g^R_{K\Sigma N}\right),
	\end{split}
\end{align}%
which can be readily obtained from those in Eq.~\eqref{su3rel}. In the SU(3)$_F$ symmetric limit, the above equations satisfy $A_1\equiv \ldots \equiv A_6\equiv B_1\equiv \ldots \equiv B_4 \equiv C_1=1$, which can be verified by inserting the coupling constants at $\kappa_{val}^{u,d}= 0.1393$ in Table~\ref{res_table}. At other quark masses, the deviations from unity represent the amount of SU(3)$_F$ breaking. Inserting the values of the coupling constants corresponding to the lowest quark mass we consider in Table~\ref{res_table} into \eqref{su3br}-\eqref{su3br3}, we find $A_1=1.045(29)$, $A_2=1.040(30)$, $A_3=1.002(25)$, $A_4=0.963(42)$, $A_5=1.017(22)$, $A_6=1.049(40)$, $B_1=1.043(28)$, $B_2=1.021(24)$, $B_3=0.990(28)$, $B_4=1.033(28)$ and $C_1=1.006(27)$, which indicate a breaking in SU(3)$_F$ by less than 10\%. Moreover, we define the average SU(3)$_F$ breaking as follows:
\begin{equation}
	\delta_{\text{SU}(3)}=\frac{1}{11} \sum_{n, X=A,B,C} \lvert 1-X_n \rvert,
\end{equation}
which amounts to $\delta_{\text{SU}(3)}=$0.014(03), 0.003(02), 0.012(02), and 0.028(17) for the quark masses at $\sim$ 150, 100, 65, and 35~MeV, respectively. This suggests for the pseudoscalar-meson couplings of the octet baryons that SU(3)$_F$ is a good symmetry in the quark-mass range we consider, which is broken by only a few percent. We have also tried a quadratic fit of $\delta_{\text{SU}(3)}$ and extracted $\delta_{\text{SU}(3)}=0.072(16)$ in the chiral limit. Fig.~\ref{su3brfig} shows the value of $\delta_{\text{SU}(3)}$ as a function of $m_\pi^2$ and the chiral extrapolations with errors. In Fig.~\ref{alpha_fig}, we plot the value of $\alpha$ as obtained from a global fit of the SU(3)$_F$ relations at each quark mass we consider. $\alpha$ slightly decreases toward the chiral limit. The deviation of $\alpha$ in the present quark-mass range is at most 10\%, whereas that of $\delta_{\text{SU}(3)}$ is less than 5\%. We infer from this that the deviation in $\alpha$ should be small in the chiral limit, as we find that the ratios of the coupling constants have weak quark-mass dependence. The SU(3)$_F$ breaking effect seems to appear in $\alpha$ rather than in SU(3)$_F$ relations ($\delta_{\text{SU}(3)}$).
We have also repeated our analysis with local source and local sink for consistency check. In Fig.~\ref{alpha_fig}, we show the values of $\alpha$ as obtained from such a setup as well, where both analysis lead to consistent results with each other.

In summary, we have evaluated the pseudoscalar-meson--octet-baryon coupling constants, $g_{\pi N\!N}$, $g_{\pi\Sigma\Sigma}$, $g_{\pi\Lambda\Sigma}$, $g_{K\Lambda N}$ and $g_{K \Sigma N} $, in two-flavor lattice QCD with the hopping parameters $\kappa_{sea},\kappa_{val}^{u,d}=$ 0.1375, 0.1390, 0.1400 and 0.1410, which correspond to quark masses of $\sim$ 150, 100, 65, and 35~MeV. The parameters representing the SU(3)$_F$ symmetry have been computed at the point where the three flavors are degenerate at the physical strange-quark mass. In particular, we have obtained $\alpha\equiv F/(F+D)=0.395(6)$, which is consistent with the prediction from SU(6) spin-flavor symmetry ($\alpha = 2/5$). The monopole mass we find leads to a $\pi N\!N$ form factor which is softer than those typically used in the phenomenological OBE potential models. The ratios of the coupling constants, which are supposed to be less prone to systematic errors, show very weak quark-mass dependence. We have discussed to what extent the SU(3)$_F$ symmetry is broken as we approach the physical masses of the $u$- and the $d$-quarks. Our results indicate for the pseudoscalar-meson couplings of the octet baryons that SU(3)$_F$ is a good symmetry, which is broken by only a few percent (at least) in the 35 MeV to 150 MeV range of the light quark masses. 

\acknowledgments
All the numerical calculations were performed on NEC SX-8R at CMC, Osaka university, on SX-8 at YITP, Kyoto University, and on TSUBAME at TITech. The unquenched gauge configurations employed in our analysis were all generated by CP-PACS collaboration~\cite{AliKhan:2001tx}. This work was supported in part by the 21st Century COE `Center for Diversity and University in Physics'', Kyoto University and Yukawa International Program for Quark-Hadron Sciences (YIPQS), by the Japanese Society for the Promotion of Science under contract number P-06327 and by KAKENHI (17070002 and 19540275).

\end{document}